# Coherent Magnetization Precession in GaMnAs induced by Ultrafast Optical Excitation

J. Qi, Y. Xu, N. Tolk

Department of Physics and Astronomy, Vanderbilt University, Nashville, TN, 37235

X. Liu, J. K. Furdyna

Department of Physics, University of Notre Dame, Notre Dame, IN 46556

I. E. Perakis

Department of Physics, University of Crete, Heraklion, Greece

We use femtosecond optical pulses to induce, control and monitor magnetization precession in ferromagnetic $Ga_{0.965}Mn_{0.035}As$. At temperatures below ~40 K we observe coherent oscillations of the local Mn spins, triggered by an ultrafast photoinduced reorientation of the in-plane easy axis. The amplitude saturation of the oscillations above a certain pump intensity indicates that the easy axis remains unchanged above ~$T_C/2$. We find that the observed magnetization precession damping (Gilbert damping) is strongly dependent on pump laser intensity, but largely independent on ambient temperature. We provide a physical interpretation of the observed light-induced collective Mn-spin relaxation and precession.

The magnetic semiconductor GaMnAs has received considerable attention in recent years, largely because of its potential role in the development of spin-based devices[1,2]. In this itinerant ferromagnet, the collective magnetic order arises from the interaction between mobile valence band holes and localized Mn spins. Therefore, the magnetic properties are sensitive to external excitations that change the carrier density and distribution. Ultrafast pump-probe magneto-optical spectroscopy is an ideal technique for controlling and characterizing the magnetization dynamics in the magnetic materials, and has been applied to the GaMnAs system by several groups[3,4]. Although optically induced precessional motion of magnetization has been studied in other magnetic systems[5], magnetization precession in ferromagnetic GaMnAs has been observed only recently[4] and has yet to be adequately understood.

In this paper, we report comprehensive temperature and photoexcitation intensity dependent measurements of photoinduced magnetization precession in $Ga_{1-x}Mn_xAs$ ($x$ = 0.035) with no externally imposed magnetic field. By comparing and contrasting the temperature and intensity dependence of the precession frequency, damping, and amplitude, we identify the importance of light-induced nonlinear effects and obtain new information on the relevant physical mechanisms. Our measurements of the photoinduced magnetization show coherent oscillations, arising from the precession of collective Mn spins. Amplitude of the magnetization precession saturates above certain pump intensity is a strong indication that direction of the magnetic easy axis remains unchanged at temperatures above about half the Curie temperature ($T_C$). The precession is explained by invoking an ultrafast change in the orientation of the in-plane easy axis, due to an impulsive change in the magnetic anisotropy induced by the laser pulse. We also find that the Gilbert damping coefficient, which characterizes the Mn-spin relaxation, depends only weakly on the ambient temperature but changes dramatically with pump intensity. Our results suggest a general model for photoinduced precessional motion and relaxation of magnetization in the GaMnAs system under compressive strain.

Time-resolved magneto-optical Kerr effect (MOKE) measurements were performed on a 300 nm thick ferromagnetic $Ga_{1-x}Mn_xAs$ ($x$ = 0.035) sample, with background hole density $p \approx 10^{20}$ cm$^{-3}$ and $T_C \approx 70$ K. The sample was grown by low temperature molecular beam epitaxy on a GaAs(001) substrate, and was therefore under compressive strain. The pump-probe experiment employed a Coherent MIRA 900 Ti:Sapphire laser, which produced ~150-fs-wide pulses in the 720 nm (1.719 eV) to 890 nm (1.393 eV) wavelength range with a repetition rate of 76 MHz. The pump beam was incident normal to the sample, while the probe was at an angle of about 30º to the surface normal. The polarization of the pump beam could be adjusted to be linear, right-circular (σ+), or left-circular (σ-) polarization. The probe beam was linearly polarized. This configuration produced a combination of polar and longitudinal MOKE, with the former dominating[6]. The temporal Kerr rotation signal was detected using a balanced photodiode bridge, in combination with a lock-in amplifier. Both pump and probe beams were focused onto the sample with a spot diameter of about 100 μm, with an intensity ratio of 15:1. The pump light typically had a pulse energy of 0.065 nJ, and a fluence of 0.85 μJ/cm$^2$.

Figure 1(a) shows our time-resolved Kerr rotation (TRKR) measurements at temperature of 20 K. The amplitude of the temporal Kerr rotation signal was found to be symmetric with respect to right or left-circularly polarized photo-excitation. In particular, we observe a superimposed oscillatory behavior at temperatures less than ~40 K, indicating magnetization precession. It is important to note that these oscillations were observed not only with σ+ and σ- polarized but also with linearly polarized pump light. The phase difference among the oscillations for σ+ or σ- pump excitation is less than ~5°. This negligible phase difference implies that the oscillatory behavior is not due to the non-thermal circular polarization-dependent carrier spin dynamics[5]. After the initial few picoseconds, where equilibration between the electronic and lattice systems occurs, the oscillations can be fitted well by the following equation (see Fig. 1(b)):

$$\theta_K(t) = A_0 \exp(-t/\tau)\cos(\omega t + \varphi) \qquad (1),$$

where $A_0$, $\omega$, $\tau$, and $\varphi$ are the amplitude, precession frequency, decay time, and initial phase of the oscillation, respectively. Some fitted parameters are shown in figure 2 as a function of pump intensity and temperature.

On a sub-picosecond time scale, the photoexcited electrons/holes scatter and thermalize with the Fermi sea via electron-electron interactions. Following this initial non-thermal temporal regime, the properties of the GaMnAs system can be characterized approximately by time-dependent carrier and lattice temperatures $T_{e/h}$ and $T_l$. Subsequently, the carrier system transfers its energy to the lattice within a few picoseconds via the electron-phonon interaction. This leads to a quasi-equilibration of $T_{e/h}$ and $T_l$, which then relax back to the equilibrium temperature via a slow (ns) thermal diffusion process. Mn precession was also observed in Ref. 4 and was attributed to the change of uniaxial anisotropy due to the increase in hole concentration[7]. In our experiment, for a typical pump intensity of 0.065 nJ/pulse, we estimate that the relative increase in hole concentration is ~0.1%. The resulting transient increase in local temperature and hole concentration leads to an impulsive change in the in-plane magnetic anisotropies and in the easy axis orientation. As a result, the effective magnetic field experienced by the Mn spins changes, thereby triggering the observed precession.

It is known that the magnetic anisotropy parameters (i.e., uniaxial anisotropy constant $K_{1u}$ and cubic anisotropy constant $K_{1c}$), which determine the direction of the easy axis in the GaMnAs system, are functions of temperature and hole concentration[4,7,8]. Thus when the GaMnAs system is excited by an optical pulse, a transient change in local hole concentration $\Delta p$ and local temperature $\Delta T$, reflecting variation of both the carrier temperature $\Delta T_{e/h}(t)$ and the lattice temperature $\Delta T_l(t)$, can lead to changes in the magnetic anisotropy parameters. Below the Curie temperature, the direction of the in-plane magnetic easy axes (given by the angle $f$) depends on the interplay between $K_{1u}$ and $-K_{1c}$. After the optical excitation, the new angle of the in-plane easy axes is given by $f(t) = f\left( \dfrac{K_{1u}(T_0 + \Delta T(t), p_0 + \Delta p(t))}{K_{1c}(T_0 + \Delta T(t), p_0 + \Delta p(t))} \right)$, where $T_0$ and $p_0$ are the initial (ambient) temperature and hole concentration[7,8]. Therefore, the in-plane easy axes may quickly assume a new direction following photoexcitation if $\Delta T(t)$ and $\Delta p(t)$ are sufficiently strong. This transient change in the magnetic easy axis, due to the change in the minimum of the magnetic free energy as function of Mn spin induced by the photoexcitation, triggers a precessional motion of the magnetization around the new effective magnetic field.

Within the mean field treatment of the *p-d* magnetic exchange interaction, the Mn spins precess around the effective magnetic field $\mathbf{H}_{eff}^{Mn}$, which is determined by the sum of the anisotropy field $\mathbf{H}_{anis}^{Mn}$ and the hole-spin mean field $JN_{hole}\mathbf{m}$. The dynamics of the hole magnetization **m** is determined by its precession around the mean field $JN_{Mn}\mathbf{M}$ due to the Mn spins, and by its rapid relaxation due to the strong spin-orbit interaction in the valence band with a rate $\Gamma_{SO}$ of the order of tens of femtoseconds[9,10]. Here, **m** (**M**) is the hole (Mn) magnetization, $J$ is the exchange constant, and $N_{hole}$ ($N_{Mn}$) is the number of holes (Mn-spins)[2]. For small fluctuations of the magnetization orientation around the easy axis, the magnetization dynamics can be described by the Landau-Lifshitz-Gilbert (LLG) equation[2], which is appropriate to apply to our experimental data at low-pump intensities (e.g., 0.065 nJ/pulse). In the adiabatic limit, where the hole spins precess and relax much faster than the Mn spins, we can eliminate the hole spins by transforming to the rotating frame[11]. In this way we obtain an effective LLG equation for the Mn magnetization **M**, whose precession is governed by the anisotropy field $\mathbf{H}_{anis}^{Mn}$ and the effective Gilbert damping coefficient including the damping $a_0$ due to e.g. spin-lattice interactions and the contribution due to the *p-d* exchange interaction, which depends on the hole concentration, the ratio of hole spin relaxation energy over exchange interaction energy, and the ratio m/M of the collective hole and Mn spins[9,10].

The LLG equation predicts an oscillatory behavior of the magnetization due to the precession of the local Mn moments around the magnetic anisotropy field $\mathbf{H}_{anis}^{Mn}(T_0+??(t), p_0+?p(t))$. The precession frequency is proportional to this anisotropy field, which is given by the gradient of magnetic free energy and is proportional to the anisotropy constants $K_{1u}$ and $K_{1c}$.[2] The magnitude of $\mathbf{H}_{anis}^{Mn}$ decreases as the ambient temperature $T_0$ or the transient temperature $DT$ increases, primarily due to the decrease in $K_{1c}$[8]. This leads to the decrease of the precession frequency as the ambient temperature or the pump intensity increases, and is consistent with the behavior observed in Fig. 2. It should be pointed out that, in the Fourier transform of each temporal signal of the oscillations, only one oscillatory mode was observed (see also in Fig. 1(b) and Fig. 1(c)). This indicates that only a single uniform-precession magnon was excited in our experiment, and the scattering among uniform-precession magnons can be neglected when interpreting damping of the Mn spin precession.

As can be seen in Fig. 2, the amplitude of the oscillations increases as the ambient temperature $T_0$ decreases or as the pump intensity increases. This result is in accord with the fact that the relative change $\boldsymbol{D}T/T_0$ and $\Delta p/p_0$, which determines the magnitude of $\boldsymbol{f}(t)$ and the photo-induced tilt in the easy axis, increase as $T_0$ decreases or as the pump intensity increases. It is important to note that in our experiment the amplitude of the oscillations saturates as the pump intensity exceeds about 4 $I_0$ ($I_0$=0.065nJ/pulse) at $T_0$=10 K. Thus, the observed saturation may indicate that the magnetic easy axis is stabilized at pump intensity larger than 4 $I_0$. We estimated that the increase of local hole concentration $\Delta p/p_0$ is about 0.4%, and the local temperature increase $\Delta T/T_0$ is about 160% using the value of specific heat of 1mJ/g/K for GaAs[12] for pump intensity ~4$I_0$. This results in the transient local temperature $T_0+\Delta T$ close to $T_C/2$. Because the magnetic easy axis is already along the [110] direction for $T_0+\Delta T$ close to or higher than $T_C/2$[8], our observed phenomenon is in agreement with the previous reported results.

Finally we turn to the oscillation damping, which is intimately related to collective localized-spin lifetimes, and consequently to spintronic device development. Figure 3 shows the fitted Gilbert damping coefficient $\boldsymbol{a}$ obtained by using the LLG equation as a function of pump intensity and ambient temperature, respectively. It can be seen that in Figure 3(a) $\boldsymbol{a}$ changes weakly with the ambient temperature and has an average value ~0.135 for a fixed pump intensity of 0.065 nJ/pulse. However, Figure 3(b) shows that $\boldsymbol{a}$ increases nonlinearly as pump intensity increases. To interpret these results, we note that, as discussed above, the *p-d* kinetic-exchange coupling between the local Mn moments and the itinerant carrier spins contributes significantly to Gilbert damping[10]. In particular, **a** increases with increasing ratio m/M. The ratio m/M is known to increase nonlinearly with increasing temperature[13] and should therefore depend nonlinearly on the photoexcitation. The Gilbert damping coefficient due to the exchange interaction also increases as hole density $p$ and hole spin relaxation rate $\frac{\Gamma_{SO}}{JMN_{Mn}}$ increase. Here, $\Gamma_{SO}$ and $\Delta p/p$ (<0.01) are relatively small. Thus we can conclude that the damping coefficient due to the *p-d* exchange interaction should increase with increasing ambient temperature ($T_0$) or increasing pump intensity ($\Delta T$ and $\Delta p$). On the other hand, we also note that damping may arise from an extrinsic inhomogeneous $\mathbf{H}_{anis}^{Mn}$ broadening attributed to a local temperature gradient due to inhomogeneities in the laser beam intensity profile and the detailed structure of the sample. This extrinsic damping effect is expected to decrease as the ambient temperature increases or the pump intensity decreases. Thus, the data in Fig. 3(a), which shows **a** only weakly dependent on ambient temperature, may result from the competition between these two mechanisms, both of which, however, predict the result in Fig. 3(b) that the damping coefficient increases nonlinearly with the increase of pump intensity. It is important to note that the LLG equation is valid only at low pump intensities. At high pump intensities, an alternative theoretical approach must be introduced[14]. So our new experimental results in the time domain are not accessible with static FMR experiments, and provide new information on the physical factors that contribute to the damping effect.

In conclusion, we have studied the photoinduced magnetization dynamics in $Ga_{0.965}Mn_{0.035}As$ by time-resolved MOKE with zero external magnetic fields. At temperatures below ~40 K, a precessional motion associated with correlated local Mn moments was observed. This precession is attributed to an ultrafast reorientation of the in-plane magnetic easy axis from an impulsive change in the magnetic anisotropy due to photoexcitation. Our results indicate that the magnetic easy axis does not change at temperatures above about $T_C/2$. We find the Gilbert damping coefficient is independent of ambient temperature but depends nonlinearly on the pump intensity, We attribute this nonlinearity to the hole-Mn spin exchange interaction and the extrinsic anisotropy field broadening due to temperature gradients in the sample. Our results show that ultrafast optical excitation provides a way to control the amplitude, precession frequency and damping of the oscillations arising from coherent localized Mn spins in the GaMnAs system.

This work was supported by ARO Grant W911NF-05-1-0436 (VU), NSF Grant DMR06-03752 (ND), and by the EU STREP program HYSWITCH (Crete).

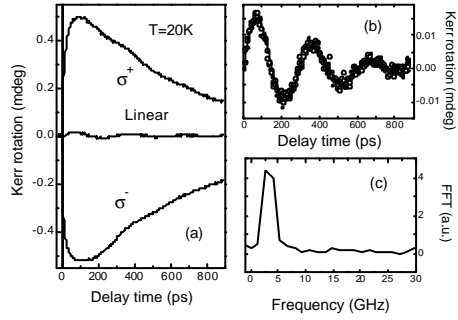

Figure 1. (a) Kerr rotation measurements for $Ga_{1-x}Mn_xAs$ ($x = 0.035$) excited by linearly-polarized and circularly-polarized light (σ+ and σ-) at a temperature of 20 K. The photon energy was 1.56 eV. Oscillations due to magnetization precession are superimposed on the decay curves. (b) Oscillation data (open circles) extracted from (a). The solid line is the fitted result. (c) Fourier transformation profile for the oscillation data in (b).

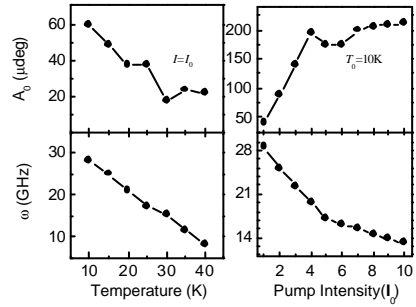

Figure 2 Amplitude $A_0$ and angular frequency $w$ as a function of temperature $T_0$ at constant pump intensity $I=I_0$; and as a function of pump intensity (in units of $I_0$) at $T_0 = 10$ K. $I_0 = 0.065$ nJ/pulse.

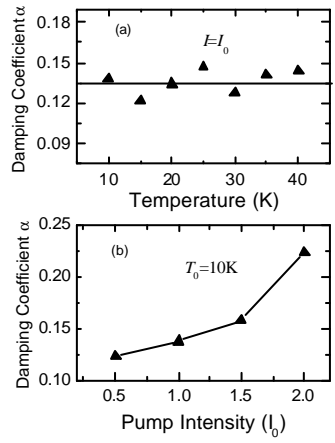

Figure 3 (a) Gilbert damping coefficient *a* as a function of temperature ($T_0$) at a constant pump intensity $I=I_0$. $I_0$=0.065 nJ/pulse; (b) Gilbert damping coefficient *a* as a function of pump intensity in units of $I_0$ at $T_0$= 10 K.